\documentclass[prl,twocolumn,aps,showpacs,floatfix,superscriptaddress]{revtex4}

\usepackage{graphicx}
\usepackage{rotating}
\usepackage{amsmath}
\usepackage{bbm}
\usepackage{subfigure}
\usepackage{color}

\renewcommand{\v}[1]{{\bf #1}}

\newcommand{\be}{\begin{equation}}
\newcommand{\ee}{\end{equation}}
\newcommand{\bea}{\begin{eqnarray}}
\newcommand{\eea}{\end{eqnarray}}

\begin{document}

\title{Is there a physical meaning of the natural orbitals? Analysis of exactly
solvable models}
\author{N. Helbig}
\affiliation{Nano-Bio Spectroscopy group and ETSF Scientific Development Centre, 
Dpto. F\'isica de Materiales, Universidad del Pa\'is Vasco, Centro de
F\'isica de Materiales CSIC-UPV/EHU-MPC and DIPC, Av. Tolosa 72, E-20018 San 
Sebasti\'an, Spain}
\author{I.V. Tokatly}
\affiliation{Nano-Bio Spectroscopy group and ETSF Scientific Development Centre, 
Dpto. F\'isica de Materiales, Universidad del Pa\'is Vasco, Centro de
F\'isica de Materiales CSIC-UPV/EHU-MPC and DIPC, Av. Tolosa 72, E-20018 San 
Sebasti\'an, Spain}
\affiliation{IKERBASQUE, Basque Foundation for Science, 48011, Bilbao, Spain}
\author{A. Rubio}
\affiliation{Nano-Bio Spectroscopy group and ETSF Scientific Development Centre, 
Dpto. F\'isica de Materiales, Universidad del Pa\'is Vasco, Centro de
F\'isica de Materiales CSIC-UPV/EHU-MPC and DIPC, Av. Tolosa 72, E-20018 San 
Sebasti\'an, Spain}
\affiliation{Fritz-Haber-Institut der Max-Planck-Gesellschaft, Berlin, Germany}

\begin{abstract}
We investigate the suitability of natural orbitals as a basis for describing
many-body excitations. We analyze to which extend the natural orbitals describe
both bound as well as ionized excited states and show that depending on the
specifics of the excited state the ground-state natural orbitals yield a good
approximation or not. We show that the success of reduced density-matrix
functional theory in describing molecular dissociation lies in the flexibility
provided by fractional occupation numbers while the role of the natural orbitals
is minor.
\end{abstract}

\date{\today}

\maketitle

Density functional theory (DFT)\cite{HK1964,KS1965} has become one of the most
widely used tools in electronic structure calculations. However, several
problems remain that cannot be adequately described with the available DFT
functionals. For one of these situations, the dissociation of small molecules,
reduced density matrix functional theory (RDMFT) has shown promising results
\cite{GU1998,GPB2005}. RDMFT is based on a one-to-one correspondence between the
ground-state one-body density matrix (1-RDM) and the ground-state many-body wave
function. This one-to-one mapping was proven by  Gilbert in 1975 \cite{G1975}.
Several functionals of the 1-RDM have appeared over the years, most of them
being functionals of the natural orbitals and occupation numbers, i.e. the
eigenfunctions and eigenvalues of the 1-RDM, rather than the 1-RDM itself 
\cite{GU1998,GPB2005,P2006,M1984,BB2002,LP2005,ML2008,RPGB2008,LHZG2009,SDLG2008,LSDEMG2009}. 

The promising results obtained with 1-RDM function- als have fueled a discussion
of the physical meaning of the natural orbitals. Due to the fractional nature of
the occupation numbers the physical significance of the natural orbitals is not
obvious. They are defined as the eigenfunctions of the 1-RDM, i.e. as purely
mathematical objects. Experience, however, shows that they are more often than
not very close to the Hartree-Fock (HF) orbitals of a system. In other words,
they seem to con- tain some physical significance as the HF orbitals are known
to do, for example from Koopman's theorem. The similarity between the HF and
natural orbitals is espe- cially striking for the homogeneous electron gas,
where the natural orbitals are plane waves due to symmetry and, hence, identical
to the HF orbitals. For the occupation numbers one obtains a smoothed step
function with reduced step size instead of the perfect step function of HF.
Generally, if one can show that the natural orbitals resemble single particle
orbitals one can connect them to single particle energies and, hence, obtain a
single-particle spectrum as an approximation to the true spectrum of a system.
Therefore, it is important to answer the question of the physical meaning
contained in the natural orbitals. Unfortunately, most RDMFT calculations
minimize the total energy which contains a small part that is only known
approximately. Consequently, one does not obtain the true natural orbitals of
the system but some approximate ones and, hence, cannot distinguish whether a
non-physical behavior of the orbital is real or just a result of the
approximation. Therefore, it is necessary to investigate the natural orbitals
for a system where one has access to the exact 1-RDM.

In this paper we investigate the natural orbitals for several model systems. We
choose one-dimensional (1D) two-electron systems because they are mathematically
identical to a one-particle system in two dimensions. Hence, the exact
two-particle wave function can be obtained numerically, and the exact 1-RDM and
natural orbitals can be calculated. As a first system we choose a single
potential well and different interaction strength between the two electrons. We
investigate the natural orbitals not only for the ground state but also for the
first excited state in order to see if two-particle excitations can be described
by only changing the occupation numbers of the ground-state natural orbitals.
The second system, two wells separated by an adjustable distance, allows us to
discuss why RDMFT yields very good results in the dissociation of molecules.
Here, we focus on the change in occupation numbers with increasing distance,
i.e. with increasing correlation. Both models allow us to smoothly increase the
correlation in the system by changing the interaction strength or the distance,
respectively. Also, they are chosen such that we can investigate both the
strongly and weakly correlated regimes in both cases. While the first model
yields a correlation between a localized and a delocalized state, the second
describes the correlation between two states localized in different parts of
space. Hence, the two models correspond to two prototypical examples of strong
correlation: Kondo physics and Hubbard correlation.

The paper is structured as follows: First, we briefly introduce the basics of
RDMFT and fix our notation. We then discuss the single well system and the
question if the natural orbitals are suitable for describing many- particle
excitations. We focus on the dissociation of a two-well system before we
conclude our findings.

\section{Theoretical Description}
The one-body density matrix of a system is calculated from its wave function via
\be\label{1rdm}
\gamma_k(\v r, \v r')=N\!\int d^3 r_2...d^3r_N
\Psi_k^*(\v r', \v r_2...\v r_N)\Psi_k(\v r, \v r_2...\v r_N),
\ee
where $\Psi_k$ is the $N$-electron wave function. For $k=0$ one obtains the
ground-state density matrix which serves as the fundamental variable in RDMFT. 
Eq. (\ref{1rdm}) can easily be modified to include spin. However, since we only 
discuss 2 electron systems in this work, the spin and spatial variables separate 
and spin is implicitly included in the symmetry of the spatial wave function. 
Instead of the 1-RDM one can employ its eigenvalues and eigenfunctions obtained
from
\be\label{defnatorb}
\int d^3r'\gamma_0(\v r, \v r')\varphi_j(\v r')=
n_j\varphi_j(\v r).
\ee 
The eigenfunctions $\varphi_j$ are known as natural orbitals with the
eigenvalues $n_j$ being their occupation numbers. The natural orbitals and
occupation numbers for excited states can be obtained in analogy to Eq.
(\ref{defnatorb}. In order for a 1-RDM to be $N$-representable the occupation
numbers have to fulfill two conditions \cite{C1963}, namely,
\be
\sum_{j=1}^\infty n_j=N, \quad 0\leq n_j\leq 1.
\ee
For non-interacting electrons the occupation numbers are strictly 0 or 1 while
for interacting electrons some, if not all, occupation numbers are fractional.
In case the system is closed-shell the natural orbitals of the two spin 
directions are identical. As a result, we can choose to work with half the
number of natural orbitals and occupation numbers between zero and two. We have
made that choice for all the closed-shell systems presented here.

Since one of the goals of RDMFT is the description of strongly correlated systems,
we also define the correlation entropy \cite{Z1995}
\be
s=-\frac{1}{N}tr(\gamma\ln\gamma)=-\frac{1}{N}\sum_{j=1}^\infty n_j\ln n_j,
\ee
where $tr$ denotes a trace. The correlation entropy describes the entanglement
of the $N-1$ variables, that were traced out in the calculation of $\gamma$, and the
remaining variable. In other words, it is a measure of the entanglement between
one particle and the other $N-1$ particles in the system. For non-interacting
particles, where the occupation numbers are strictly zero and one only, the
correlation entropy is zero. A maximum contribution is obtained for $n_j=e^{-1}$
but the case where all occupation numbers obtain this value is usually not
compatible with the total number of particles. Hence, as we will see in the
following, the signature of strong correlation is half-occupation for the
natural orbitals. For closed-shell systems, if one chooses to work with
occupation numbers between zero and two, each $n_j$ needs to be divided by two
and the whole sum multiplied by two, for the spin summation, in order to obtain 
the correct correlation entropy. 

We consider 1D two-electron systems, hence, the Hamiltonian is given by (atomic 
units are used throughout)
\be\label{hamiltonian}
-\frac{d^2}{2dx^2_1}-\frac{d^2}{2dx^2_2}
+v_{\mathrm{ext}}(x_1)+v_{\mathrm{ext}}(x_2)+v_{\mathrm{int}}(|x_1-x_2|),
\ee
where $v_{\mathrm{ext}}$ denotes the external potential and $v_{\mathrm{int}}$
the electron-electron interaction. As we can see, the Hamiltonian is
mathematically equivalent to a single electron in two-dimensions with the 2D
external potential 
\be
v_{\mathrm{ext}}^{\mathrm{2D}}(x_1,x_2)=
v_{\mathrm{ext}}(x_1)+v_{\mathrm{ext}}(x_2)+v_{\mathrm{int}}(|x_1-x_2|).
\ee
The wave function for this problem can be calculated with any numerical code
that can treat non-interacting electrons in 2D. The 1-RDM and the natural
orbitals and occupation numbers can then be obtained via Eqs. (\ref{1rdm}) and
(\ref{defnatorb}). We use the \texttt{OCTOPUS} code 
\cite{MCBR2003,CAORALMGR2006} for all the calculations presented here. The 
calculations were performed in a finite box with zero boundary conditions. Also, 
we consider systems at zero temperature as reflected in the Hamiltonian Eq.
(\ref{hamiltonian}).

\section{Description of excitations}

We consider a single potential well of hyperbolic cosine form, i.e. the external
potential is given by
\be\label{singlecosh}
v_{\mathrm{ext}}(x)=-\frac{v}{\cosh^2(\kappa x)}.
\ee
The single particle eigenvalues of the system are given by \cite{LL1977}
\be
\epsilon_n=-\frac{\kappa^2}{8}\left[\sqrt{1+\frac{8v}{\kappa^2}}-1-2n\right]^2
\ee
where it is understood that the square bracket needs to be positive. Hence, for 
$v/\kappa^2<1$ there exist only a single bound state while for $1<v/\kappa^2<3$
there are exactly two bound states for non-interacting particles. In our
discussion of the physical interpretation of the natural orbitals we consider
two scenarios: In both cases we choose $\kappa=1$, but $v = 0.9$ in the first 
and $v = 2.0$ in the second case. While the situation with exactly one bound 
state seems rather artificial for quantum chemistry it is frequently encountered 
in semiconductor nanostructures or in metals in the context of the Kondo effect 
and Anderson impurities, to name two examples. For the electron-electron 
interaction we choose a finite-range interaction, namely,
\be\label{vint}
v_{\mathrm{int}}(x)=-\frac{b}{\cosh^2(x_1-x_2)}
\ee
with a variable interaction strength $b$. The external potential, Eq.
(\ref{singlecosh}) is symmetric and, therefore, the two-particle wave function
can be chosen as an eigenfunction of the parity operator. As a result, the 1-RDM
is symmetric under parity and the natural orbitals are simultanious
eigenfunctions of the 1-RDM and the parity operator. Consequently, we can order
them with increasing number of nodes, i.e. starting with a nodeless symmetric
natural orbital followed by an antisymmetric one with one node and so on.
Alternatively, one can order the natural orbitals with decreasing occupation,
i.e the first natural orbital is the one with the highest occupation. The two
orders do not necessarily coincide, as one of our examples shows. Throughout
this paper we have chosen to order the natural orbitals according to their
occupation number.

If there is only one bound state, the two non-inter\-acting electrons occupy this
state in a singlet configuration. However, if we continuously increase the
interaction between the two electrons the energy of this state increases and
eventually the two-electron wave function resembles one electron in the bound
state and the other in an extended state, i.e. the system ionizes. If the system
contains two bound states then the increase in the interaction strength also
leads to an increase in the energy of the non-interacting ground-state
configuration, however, the new ground-state is still localized, i.e. the 
system does not ionize due to the existence of the second bound state. Of course,
upon further increase of the interaction this system also ionizes but only at an 
interaction strength significantly higher than for the system with only one 
bound state. In the following we discuss the behavior of the natural orbitals 
for different interaction strength $b$. 

In order to answer the question whether an excitation of the many-body system
can be described by the natural orbitals of the ground state we investigate them
for the ground- and the first excited states. If the natural orbitals of these
two states are indeed very similar then the excitation can be described by a
change in the occupation numbers alone. As a result, the energy spectrum of the
many-body system could be obtained from assigning single-particle energies to the
natural orbitals and calculating the appropriate weighted sum to obtain the
many-body energy. All calculations in this section were performed in a box 
ranging from -15 to 15 with a spacing of 0.05. 

\begin{widetext}

\begin{figure}
\includegraphics[width=0.95\textwidth,clip]{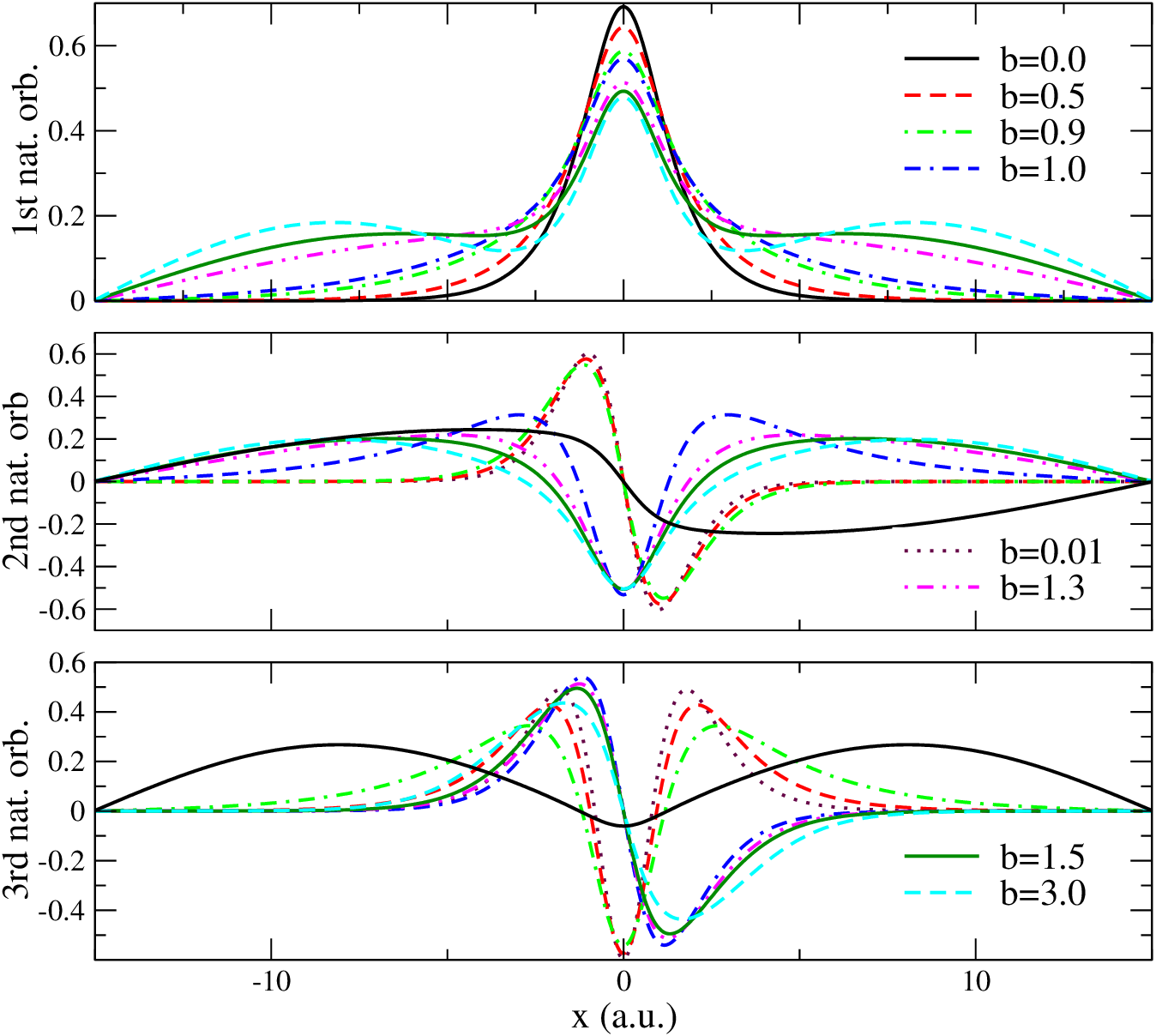}
\caption{\label{phigs} First, second, and third natural orbital of the 
ground-state density matrix for different interaction strength. For 
non-interacting electrons the second and third natural orbital are extended but
immediately localize when the interaction is turned on.}
\end{figure}

\end{widetext}

\begin{figure}
\includegraphics[width=0.45\textwidth,clip]{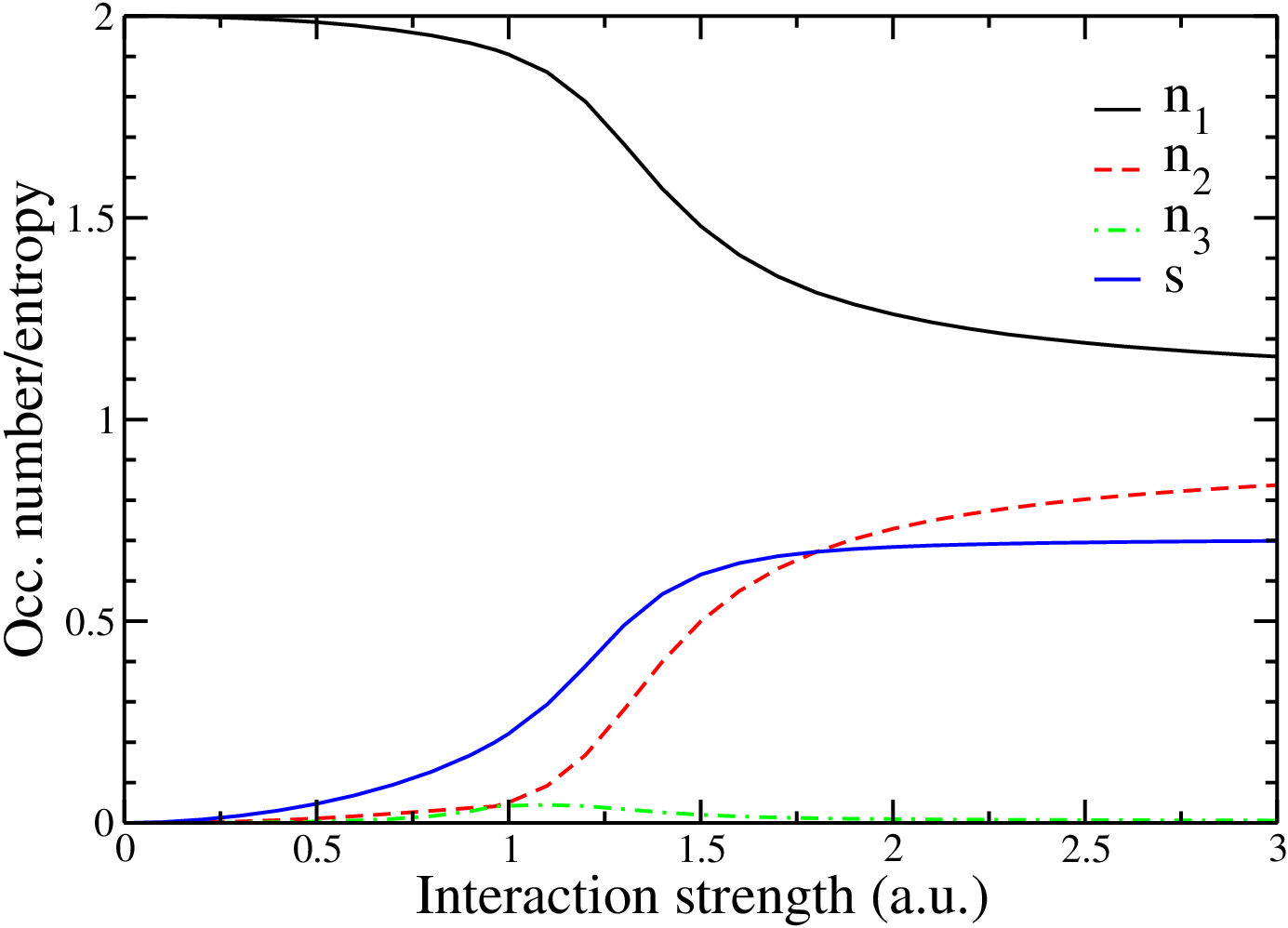}
\caption{\label{occnumbgs}Three largest occupation numbers of the ground-state 
density matrix and correlation entropy, $s$, as a function of the interaction strength 
$b$.}
\end{figure}

Fig. \ref{phigs} shows the first three natural orbitals of the ground state for
an external potential with $v=0.9$. For non-interacting electrons, i.e. $b=0.0$,
only the first orbital is occupied while all other orbitals are empty. The empty 
orbitals are numerically not accessible from a diagonalization of the density 
matrix and, since they are all degenerate with occupation number zero, they are 
only defined up to unitary transformations in the degenerate subspace. The most 
natural choice is, of course, the single particle eigenstates of the problem, 
which is what we plotted in Fig. \ref{phigs} for $b=0.0$. As we can see, the 
orbitals are essentially the eigenstates of the box, i.e. $\cos$ and $\sin$ functions, 
with a small modification at the position of the potential well. A tiny 
interaction, $b=0.01$, leaves the first natural orbital unchanged but, as we can 
see in Fig. \ref{phigs}, the higher natural orbitals become very localized.
This holds true for all the natural orbitals that we included in our calculation, 
not only these two. At first this is surprising since the small interaction 
represents a small perturbation of the non-interacting system. However, for the 
natural orbitals this perturbation acts on a highly degenerate subspace, all 
natural orbitals previously had zero occupation. Hence, all the extended states
mix and form localized orbitals. Consequently, even an infinitesimal interaction 
leads to the whole set of natural orbitals to localize in the potential well. 
Since the occupation number of the first natural orbital is still almost 2
and all remaining occupation numbers are very small, see Fig. \ref{occnumbgs}, 
the 1-RDM is almost identical to the one for non-interacting electrons, 
confirming that the perturbation is indeed small.

From Fig. \ref{phigs} we see that around $b=0.9$ the shape of the 2nd natural
orbital changes quite dramatically from an orbitals with one node to one with
two nodes. However, a comparison with the third natural orbital reveals that
this natural orbital undergoes the opposite transition. Hence, it seems that the
two orbitals have switched places which is confirmed by a look at the occupation
numbers in Fig. \ref{occnumbgs} which are identical for $b$ slightly above 0.9.
In other words, the drastic changes in the second and third natural orbitals are
a result of the fact that we always order them with decreasing occupation
number. If we had ordered them according to their parity, each natural orbital
would show a smooth change along the whole range of $b$ and the occupation
numbers in Fig. \ref{occnumbgs} would cross.

All three natural orbitals show an increasing delocalized part above $b=1.0$.
Again, since we are running the simulations in a finite box, these extended
states resemble the eigenstates of the box. We have ensured that the box is
large enough for the localized parts of the orbitals to remain unaffected by a
change in the size of the box. However, the extended parts of course depend on
the choice of box size. We interpret these results as an increase in the degree
of ionization of the ground state of the system. In other words, the two
electrons do not occupy the same single-particle state, i.e. the one bound state
of the system, anymore. Due to the increased interaction one of the electrons is
forced to occupy a different level, and since no other bound state is available
it occupies an extended state. Of course, this only approximately describes the
situation since the two electrons are interacting and, hence, the notion of
single-particle levels is not totally appropriate. The occupation numbers, Fig.
\ref{occnumbgs}, show the increase in interaction by deviating more and more
from their non-interacting values, zero and two. For $b\rightarrow\infty$ the
two largest occupation numbers approach one and are degenerate. Looking at the
evolution of the first two natural orbitals for $b>1.0$ it is clear that one can
form a linear combination with one natural orbital completely localized while
the other represents a slightly modified box states. Physically this describes
the situation at infinite interaction strength where the ground state contains
one localized and one completely delocalized electron. We also note that the two
degenerate natural orbitals are both of even parity. Fig. \ref{occnumbgs} also
shows that the correlation entropy increases with increasing interaction
strength and converges to a value of about 0.7. In other words, for large
interaction the situation closely resembles what is usually described as Kondo
physics: the strong correlation between a localized and a delocalized state.

\begin{widetext}

\begin{figure}
\includegraphics[width=0.98\textwidth,clip]{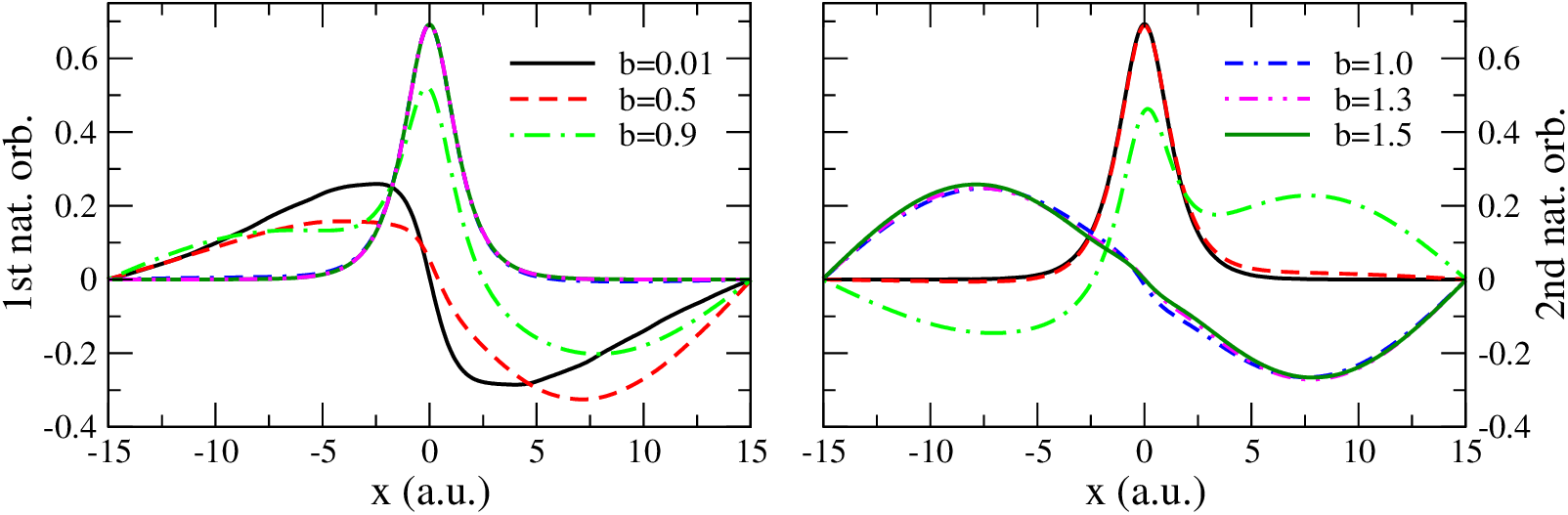}
\caption{\label{phiex} First and second natural orbital of the density matrix 
of the first excited state, a spin triplet, for different interaction strength. 
For $b = 0.0$ the first two excited 2-particle states are degenerate leading to 
an ambiguity in the natural orbitals. A small interaction, $b = 0.01$ lifts the 
degeneracy and yields the results shown here.}
\end{figure}

\end{widetext}
The first two natural orbitals for the first excited state of our two-electron
system are plotted in Fig. \ref{phiex}. Since the first excited state is a spin
triplet the excitation con- tains a spin flip of one of the two electrons. As we
can see, the first natural orbitals is delocalized for non-interacting electrons
but becomes localized between $b=0.9$ and $b=1.0$. The second natural orbital
shows the opposite trend, and again, this change is due to a change in the
ordering of the natural orbitals as the two occupation numbers become identical
and change order. One striking result, however, is the fact that over the whole
range of interactions, one of the two natural orbitals is delocalized. This is
not surprising as we expect the first excited state of a system with only one
bound state to be partially ionized. As a result, this excitation cannot be
described by changing the occupation of the ground-state natural orbitals at
least in the range of $b<1.0$, where all ground-state natural orbitals are very
well localized. In other words, while a description of many-body excitations via
single particle excitations is no problem for non-interacting electrons, for a
small interaction the natural orbitals are not suited for such a description. We
note that for non-interacting electrons the first and second excited 2-particle
states are degenerate. As a result, any linear combination of the two states is
an eigenfunction of the Hamiltonian leading to an ambiguity of the natural
orbitals for the first excited state. We avoid this effect by plotting the
natural orbitals for a very small interaction of $b=0.01$ which is sufficient to
lift the degeneracy.

The inability of the natural orbitals to describe the excitation above is likely
a result of the fact that our system has only one bound state and, hence, any
excitation involves an ionization. Therefore, we increase the depth of our well
by choosing $v=2.0$ which results in a second bound state for the
non-interacting electrons.

\begin{widetext}

\begin{figure}
\includegraphics[width=0.95\textwidth,clip]{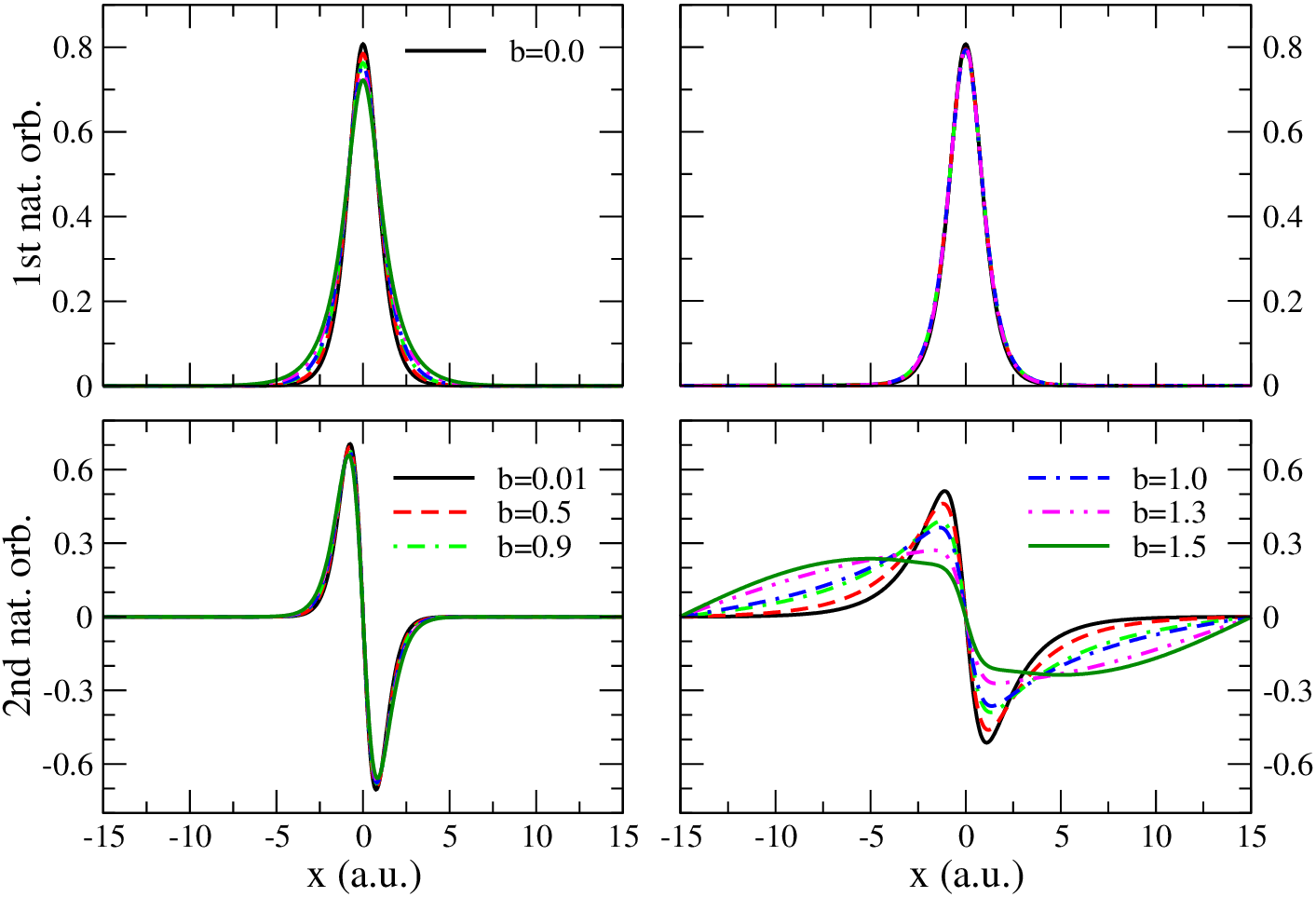}
\caption{\label{phiv2} First and second natural orbitals of the ground-state 
density matrix (left) and the density matrix of the first excited state (right) 
for different interaction strength for a potential with $v = 2.0$. The ground 
statae is a spin singlet while the first excited state is a triplet. All graphs 
share the same color coding unless otherwise stated. For $b = 0.0$ the second 
natural orbital of the ground state is numerically ill defined. Also, the same 
degeneracy of the first two excited states as before is encountered for 
non-interacting electrons.}
\end{figure}

\end{widetext}

Fig. \ref{phiv2} shows the first and second natural orbital of the ground- and
first excited states of the system with $v=2.0$. As we can see, over the whole
range of interaction the natural orbitals of the ground-state hardly change at
all. Due to the much deeper well the first two natural orbitals stay well 
localized within the well even for relatively large interaction strength. Also, 
the change in the first natural orbital of the first excited state is almost 
unnoticeable on the scale of the plot. The second natural orbital, however, 
shows quite a pronounced change with increasing interaction strength. It becomes 
increasingly more delocalized with larger interaction strength and obtains 
features of the box state similar to the behavior of the natural orbitals of the 
ground state for the smaller potential well, see Fig. \ref{phigs}. This can be 
understood if one keeps in mind that the energy of the excited state is closer 
to the continuum and, therefore, acquires a certain degree of ionization much 
earlier than the ground state. We also observe that the form of the two natural 
orbitals, especially for small interaction strength, is similar for the ground- 
and the first excited state. Looking at the occupation numbers we notice that 
the first two orbitals for the excited state are equally occupied with an 
occupation of $1.000$ at $b = 0.01$ decreasing to $0.998$ at $b = 1.5$. For the 
ground state the occupation of the first natural orbital decreases from $2.000$ 
to $1.957$, respectively. Consequently, removing one electron from the first 
natural orbital of the ground state and placing it into the second one yields a very good description of
the first excited state for small interaction strength and is expected to still 
be a reasonably good approximation for intermediate $b$. 

Our calculations suggest that excitations between many-body states that are of 
similar nature, for example two well-localized bound states, can be described by
only changing the occupation numbers of the ground-state natural orbitals. Any 
excitations that involve an ionization of the system, however, cannot be 
expected to be well described by using the ground-state natural orbitals. Also, 
an excitation from a localized low-energy state to a high-energy Rydberg state 
in an atom cannot be expected to be well described. In order to obtain the 
excitation energies from the natural orbitals one needs to assign energy levels 
to the natural orbitals, which is a challenge of its own as the natural orbitals 
are not defined as the eigenstates of a Hamiltonian.

\section{Molecular dissociation}\label{dissociation}

In order to investigate why RDMFT is very successful in describing molecular
dissociation, we again use a 1D model system. The external potential now
consistes of two wells and is given by
\be
v_{\mathrm{ext}}(x)=-\frac{v}{\cosh^2(x-d/2)}-\frac{v}{\cosh^2(x+d/2)},
\ee
where $d$ describes the distance between the two wells. We choose $v=0.9$ for
all the calculations in this section. The interaction remains of the form Eq.
(\ref{vint}) with $b=0.5$. The calculations are performed in a box ranging from -20
to 20 with a grid spacing of 0.05. The model resembles a diatomic molecule and
therefore dissociates into two independent single-well fractions. It was shown
that within DFT the independence of the two fragments is ensured by the
appearance of a peak at the mid-point of the exact Kohn-Sham (KS) potential
\cite{P1985, BBS1989, GRB1995, GB1996,HTR2009}. However, none of the commonly
used DFT functionals reproduces the exact behavior which, at least partially,
explains the failure of DFT in describing molecular dissociation. Within RDMFT
the dissociation of small molecules is very well described by even the first
generation functionals \cite{GU1998,GPB2005} and further improved by the more
recent ones \cite{GPB2005}. Hence, the question arises why RDMFT is so much more
successfull in this case. Either the natural orbitals are more suited to
describe molecular dissociation than the KS orbitals or the additional freedom
of fractional occupation numbers makes the difference.

\begin{widetext}

\begin{figure}
\includegraphics[width=0.95\textwidth,clip]{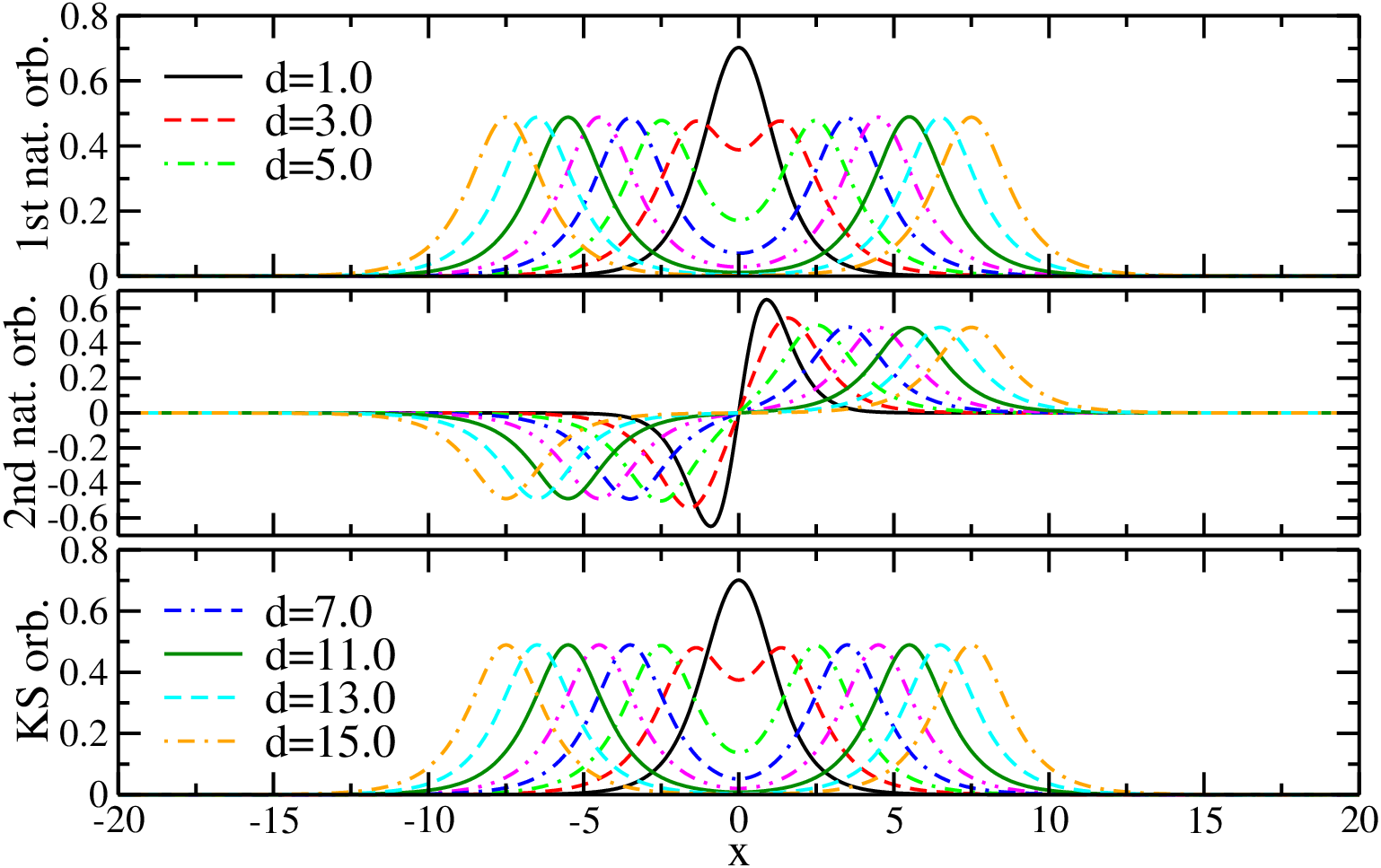}
\caption{\label{phi} First and second natural orbital as a function of distance 
between the two wells. For comparison the single occupied KS orbital is also
shown. (All graphs are in a.u.)}
\end{figure}

\end{widetext}

\begin{figure}
\includegraphics[width=0.45\textwidth,clip]{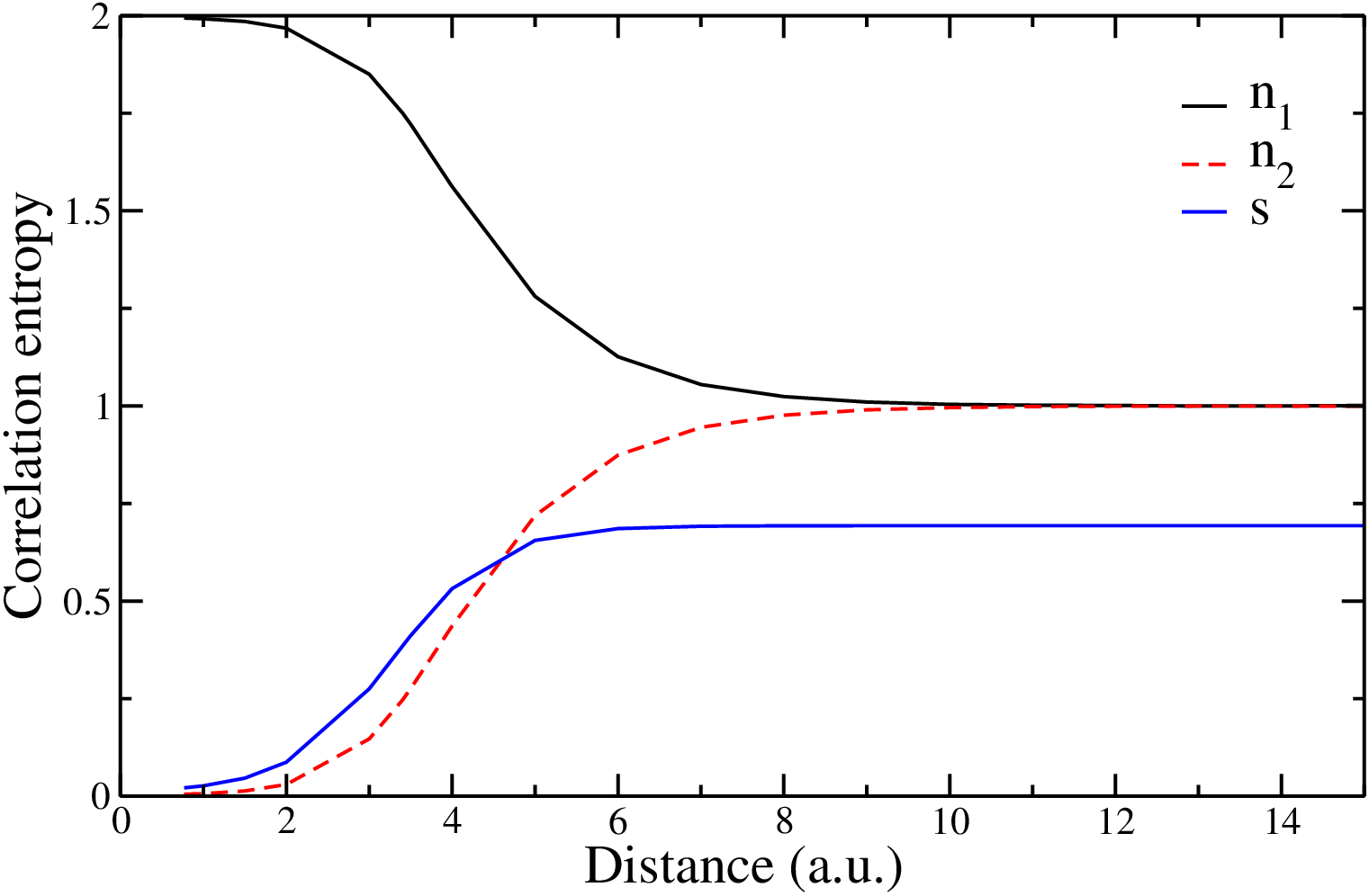}
\caption{\label{entropy}Occupation numbers and correlation entropy, s, of the
ground-state density matrix as a function of distance between the two wells.}
\end{figure}

Fig. \ref{phi} shows the first and second natural orbitals for different
distances $d$ between the wells. For comparison we also included the doubly
occupied first KS orbital. We emphasize that this is the exact KS orbital which,
for two electron singlet systems, is given as $\sqrt{n(\v r)/2}$. The occupied 
KS orbital is very similar to the first natural orbital. However, while the KS orbital is doubly occupied for all distances the
occupation of the first natural orbital changes from almost two at small
distances to about one for distances larger than 10 a.u., see Fig. 
\ref{entropy}. At the same time the occupation of the second natural orbital
increases from zero to one such that both natural orbitals are half occupied at
large distances (for closed-shell system an occupation of one corresponds to
half occupation). Strictly speaking the occupation numbers of these two orbitals
only become degenerate in the limit $d\rightarrow \infty$. Numerically, however,
we observe that they are identical for distances larger than 10 a.u. As a
result, the numerical calculation can produce natural orbitals that violate
parity, for example if one works with an asymmetric grid. At infinite separation
the degeneracy of the first two natural orbitals implies that one can use any
linear combination of these orbitals without changing the 1-RDM. Especially, one
can choose one natural orbital to be localized at each well mimicking the
situation that appears in nature due to the small parity violating fluctuations
that are present there.

Fig. \ref{entropy} also shows that at small distances the first natural orbital
is almost fully occupied while the second is nearly empty. All other occupation
numbers are too small to be visible on the scale of this figure. The fact that
with increasing distance the natural orbitals become half occupied implies that
the amount of correlation in the system increases with increasing distance. To
quantify the correlation we calculate the correlation entropy as a function of
$d$. As expected the entropy increases
with increasing distance and saturates at a value of $s\approx 0.69$, the
equivalent of four orbitals being occupied by half an electron, at around
$d=10$ a.u. the distance where the occupation numbers of the first two natural
orbitals become degenerate.

\section{Conclusions}

We investigated the behavior of the natural orbitals in two different
situations, a single well and two wells with different distances. It was shown 
that the natural orbitals of the single well system all localize if the 
electrons are interacting, even if the interaction strength is tiny. As a result, 
the ground-state natural orbitals are not a good description of an excited state 
if the latter is partially ionized. On the other hand, the natural orbitals of 
a bound excited state are rather similar to the ground-state ones such that a 
description of the excitation by a change in the occupation numbers alone is a 
good approximation. It was also shown that an increase in interaction strength 
leads to a partial ionization of the states which manifests itself in a 
partially delocalized character of the natural orbitals.

The reason RDMFT is highly successful in the description of molecular 
dissociation, and most likely also other strongly correlated situations, lies 
in the freedom of fractional occupation numbers. At all distances the first 
natural orbital closely resembles the occupied KS orbital of DFT. At large 
distances, the occupied KS orbital remains symmetric with equal contributions 
at each well. In contrast to that, in RDMFT it is possible to occupy two natural 
orbitals with the same fraction of a particle. In other words, two natural 
orbitals can become degenerate with respect to their occupation number and, 
therefore, one can perform unitary transformations in the degenerate subspace 
without changing the one-body density matrix. Therefore, in the dissociation 
limit one can choose the two natural orbitals such that they each localize at 
one of the fragments in resemblance of the behavior of the electrons in reality.

In the future, the problem of assigning energy levels to the natural orbitals 
needs to be addressed. It was shown that the natural orbitals can be obtained 
as the eigenstates of a single-particle Hamiltonian with a non-local
external potential \cite{G1975,P2005}. However, this Hamiltonian is highly 
non-unique. More specifically, its eigenvalues are undetermined meaning that 
the energy levels corresponding to the natural orbitals are arbitrary. A 
Hamiltonian with a local potential is not guaranteed to exist for an arbitrary 
set of natural orbitals but it might provide a good first approximation. Work 
in this direction is currently in progress.

We acknowledge funding by the Spanish MEC (FIS2007-65702-C02-01), 
``Grupos Consolidados UPV/EHU del Gobierno Vasco'' (IT-319-07), the
European Community through e-I3 ETSF project (Contract Number 211956).

\end{document}